# WS2-graphite dual-ion battery

Sebastiano Bellani,[†] Faxing Wang,[‡] Gianluca Longoni,[†] Leyla Najafi,[†] Reinier Oropesa-Nuñez,[∇] Antonio E. Del Rio Castillo,[†] Mirko Prato,[#] Xiaodong Zhuang,[‡] Vittorio Pellegrini,[†,∇] Xinliang Feng[‡] and Francesco Bonaccorso*[†,∇]

[†] Graphene Labs, Istituto Italiano di Tecnologia, via Morego 30, 16163 Genova, Italy

[‡] Center for Advancing Electronics Dresden (cfaed), Chair for Molecular Functional Materials, Department of Chemistry and Food Chemistry, Technische Universität Dresden, Mommsenstrasse 4, 01062 Dresden, Germany

[∇] BeDimensional Srl, via Albisola 121, 16163, Genova, Italy

[#] Materials Characterization Facility, Istituto Italiano di Tecnologia, via Morego 30, 16163 Genova, Italy

**Abstract**

A novel $WS_2$-graphite dual-ion battery (DIB) is developed by combining together a conventional graphite cathode and high-capacity few-layer $WS_2$ flakes anode. The $WS_2$ flakes are produced by exploiting wet-jet milling (WJM) exfoliation, which allows mass production of few-layer $WS_2$ flakes in dispersion, with an exfoliation yield of 100%. The $WS_2$-anodes enable DIBs, based on hexafluorophosphate ($PF_6^-$) and lithium ($Li^+$) ions, to achieve charge specific capacities of 457, 438, 421, 403, 295 and 169 mAh $g^{-1}$ at current rates of 0.1, 0.2, 0.3, 0.4, 0.8 and 1.0 A $g^{-1}$, respectively, outperforming conventional DIBs. The $WS_2$-based DIBs operate in the 0 to 4 V cell voltage range, thus extending the operating voltage window of conventional $WS_2$-based Li-ion batteries (LIBs). These results demonstrate a new route towards the exploitation of $WS_2$, and possibly other transition metal dichalcogenides (TMDs), for the development of next-generation energy storage devices.

Today more than ever, electrical energy storage (EES) technologies are playing a pivotal role in transportation[1,2] and on-grid application.[3] Moreover, advanced EES systems are considered for new-generation grids to effectively face the energy harvesting from intermittent renewable sources and the energy production/deployment spike leveling.[4,5,6] Among the EES systems, Li-ion batteries (LIBs)[7,8,9] are used in high-end mobile electronics[1,10] and electric vehicles.[1,11] Furthermore, their integration in small-to-mid size stationary storage is believed to be a valid alternative to traditional technologies, including lead-acid battery,[12,13] flywheels[14] and small hydropower system.[15] However, the energy/power density and energy cost requirements for future electric vehicles and stationary storage are becoming stringent, as these applications are hitting the mass market.[11] In particular, volumetric and gravimetric energy densities superior to 300 Wh L$^{-1}$ and 250 Wh kg$^{-1}$, respectively, and a cost inferior to \$130/kWh are pursued in the manufacturing of future full-electric vehicles.[1,2,16,17] In the meanwhile, stationary applications, referred to kW/MW/GW worth systems will require battery manufacturing cost limited to \$100s/kWh and a cycle-ability extended to thousands of charge-discharge (CD) cycles.[18] These challenges are progressively expanding the research on LIBs alternatives,[19] including Na-ion[20] and K-ion[21], which could address the concerns for Li abundance and costs,[20,21] or Li-S[22,23] and Li-air batteries[22,24,25], which are promising for their use in ultrahigh energy density systems.[22,24] In this context, dual-ion batteries (DIBs) represent a novel electrochemical energy storage architecture.[26,27,28] These batteries are able to operate across a wider voltage window (above 4 V, depending on the materials and electrolytes used)[29] compared to the ones based on common chemistries (*i.e.*, ~3.3 V/3.6 V/3.7 V for LiFePO$_4$-/LiCoO$_2$-/LiMn$_2$O$_4$-based commercial LIBs).[1,7,8,9] Moreover, differently from standard LIBs, which are based on the so-called "rocking chair" mechanism (*i.e.*, the migration of metal cations between cathode and anode host materials during charge/discharge (CD) process),[7,8,9] in the typical CD process of a DIB, the electrochemically stable cations (Li$^+$, Na$^+$, K$^+$)[29,30,31,32,33,34,35,36,37] and anions (typically polyatomic, such as PF$_6^-$, ClO$_4^-$, AsF$_6^-$ and SbF$_6^-$)[29,30,31,32,33,34,35,36,37] are simultaneously intercalated/de-intercalated into/from anode and cathode, respectively. Experimentally, only a few materials have been reported to host anions reversibly, and thus being able to work as suitable cathodes for DIBs. In fact, the cathode material must have interstitial sites to accommodate the anions, while positive electric charges are transferred to the host lattice *via* the external circuit.[38,39,40] For these reasons, graphite, being formed by stacked graphene layers with an interlayer *d* spacing of ~3.35 Å[41], is the most consolidated cathode for DIBs.[26,27,28] Recently,

"*nongraphitic chemistries*" (*e.g.*, redox-active metal-organic frameworks,[38] polycyclic aromatic hydrocarbon molecular solid,[40] nitrogen-containing organic host lattice[42] and organic-derived nitroxide radicals[43,44]) have also been proposed to increase the CD reversibility, which is negatively affected by the decomposition of conventional electrolyte solvent at the anion-intercalating potentials of graphite (> 4.5 V *vs.* $Li^+/Li^{38}$).[29,30,31,32,33,34,35,36,37] *Vice versa*, any anode materials suitable for metal-ion batteries can be, in principle, adopted for DIBs.[26,27,28] Typically, graphite is also used as anode in DIBs due to its low cation-intercalating potential (< 1.0 V *vs.* $Li^+/Li$).[45,46,47] The overall DIB cell acquires in this way a graphite-based symmetric configuration, whose ease of assembling is a major advantage. Recently, novel metal-graphite DIBs have started to exploit metal foils (*e.g.*, Al, Sn, Pb, K, Na, Si and Sb) as both anodes and current collectors,[27,34,48,49] defining new battery designs.

For the sake of enlarging the class of anode materials for DIBs, in this work, we first report a novel DIB based on $WS_2$, which is representative member of the wide class of transition metal dichalcogenides (TMDs).[50,51] $WS_2$ has also been proven to be more chemically stable (against oxidation and high-rate thermal decomposition) compared to the most investigated 2D material, *i.e.*, $MoS_2$.[52,53,54] Moreover, *ab initio* theory predicts that, among all the S-based TMDs, $WS_2$ shows the highest electrical mobility due to its reduced electron/hole effective mass (~0.30.4).[55,56] Experimentally, transport measurement reported room-temperature charge carrier mobilities of liquid-gated single- and double-layer $WS_2$ as high as ~50 $cm^2V^{-1}s^{-1}$.[57,58] Lastly, $WS_2$, as well as other TMDs, have been successfully exploited as negative electrodes in various ion-battery system.[59,60,61] In fact, they possess a layered structure held together by van der Waals interactions that ensures the space for efficient cation intercalation/de-intercalation.[62,63] In particular, the interlayer spacing of $WS_2$ (> 6 Å[64,65] for both its thermodynamically stable 2H and 3R phases[66,67] belonging to space group $P6_3/mmc$[64] and $R3m$[64], respectively) is significantly larger compared to graphite (3.35 Å).[41] This means that the intercalation/de-intercalation processes of $Li^+$ for layered $WS_2$ is facilitated compared to graphite. According to the sequential lithiation reactions of $WS_2$ (*i.e.*, $WS_2 + xLi^+ + xe^- \rightarrow Li_xWS_2$ at ~1.1 V vs. $Li^+/Li$ followed by $Li_xWS_2 + (4-x)Li^+ + (4-x)e^- \rightarrow W + 2Li_2S$ at ~0.6 V vs. $Li^+/Li$, $0 \leq x \leq 1$),[68,69,70,71,72,73,74] the theoretical specific capacity of $WS_2$ results to be 433 mAh $g^{-1}$ for the uptake of 4 $Li^+$ for unit formula.[68,69,70,71,72,73,74] This value is higher than the theoretical reversible specific capacity of graphite anodes (372 mAh $g^{-1}$ for the end-compound $LiC_6$),[75,76,77] for which Li storage is limited by sites within a $sp^2$ hexagonal carbon structure.[75,76,77]

In addition, further de-lithiation of the electrode leads to the formation of elemental S domains, while metallic W is maintained as an electrochemically inert buffer.[68,69,70,71,72,73,74] The co-existence of these two elements at the discharged state of the electrode entails a dual advantage. Firstly, metallic W buffer positively affect the overall electronic conductivity in the electrode and secondly, elemental S, formed after the first $WS_2$ lithiation/de-lithiation loop,[68,69,70,71,72,73,74] participates to further charge/discharge loops that are relevant for DIBs. These latter loops guarantees, in principle, a higher, compared to the first lithiation reactions of $WS_2$, theoretical specific capacity of 1675 mAh $g^{-1}$, according to the reaction $S + 2Li^+ + 2e^- \rightarrow Li_2S$ occurring at ~2.2 V vs. $Li/Li^+$.[68,69,70,71,72,73,74,78,79,80] The last process, commonly exploited on the cathode side in Li-S batteries,[81] is exploited on the anode side of DIBs, despite the high operational voltage of the lithiation-de-lithiation process of S.[78,79,80] This issue of high operation voltage is solved in DIBs,[26,27,28] since the anion intercalating graphitic cathode operates at higher potentials *vs.* $Li^+/Li$.[26,27,28] Energy density outputs of DIB systems can therefore offer satisfactory results in term of capacity while effectively eliminating bottlenecks of typical Li-S battery configuration such as metallic Li protection and $Li_2S$ dissolution.[81] Finally, TMD-based electrodes express significantly less volumetric expansion upon lithiation (~100% expansion for conversion reaction of TMD to $LiS_2$ and metal)[45,46,47] compared to the one experienced by graphite or other recently investigated anode materials (up to 400% for $Si^{[82,83]}$).[59,60,61]

We herein propose high-pressure wet-jet milling (WJM) exfoliation of $WS_2$ powder yielding $WS_2$ flakes having average lateral size of ~400nm and thickness of ~1.5nm exploiting a production approach that is capable to bridge the gap between laboratory-scale studies and commercial applications. We demonstrate that the WJM exfoliation[84] allows for large-scale and free-material loss production (*i.e.,* volume up to 8 L $h^{-1}$ at concentration of 10 g $L^{-1}$ and exfoliation yield of 100%)[84] of few-layer $WS_2$ flakes in dispersion. The as produced few-layer $WS_2$ flakes are used as anodes in DIBs, achieving reversible specific capacities of 457, 438, 421, 403, 295 and 169 mAh $g^{-1}$ at the current rates of 0.1, 0.2, 0.3, 0.4, 0.8 and 1 A $g^{-1}$, respectively. The exploitation of few-layer $WS_2$ flakes as anode, coupled with graphite-based cathode, allows for the realization of DIBs operating in the 0-4 V range, with an average value of ~2.4 V. The operating cell voltage is remarkably superior to that expressed by the typical TMD-based LIBs (< 2V), opening the way toward the exploitation of TMDs as low-cost and high-capacity anode materials for novel EES.

**Production and characterization of $WS_2$ flakes.** As schematically illustrated in **Figure 1**, the WJM exfoliation process comprises a first step of preparing a dispersion of $WS_2$ powder with N-methyl-2-pyrrolidone (NMP) as dispersant solvent, and a subsequent step of exfoliation of the dispersed $WS_2$ flakes during their exposure to the hydrodynamic forces generated through high-pressure (250 MPa) compression of the dispersant fluid phase, as applied by an hydraulic piston.[84] Afterward, the sample is cooled down in form of a liquid dispersion by means of a chiller.

Additional details are reported in Experimental Section and Supporting Information (**Figure S1**), as well as in our recent work.[84] In order to evaluate the effectiveness of the WJM exfoliation of $WS_2$, it is useful to define the following of Figures of Merit (FoM): 1) the time required to obtain 1 g of exfoliated $WS_2$ powder in dispersion after the exfoliation process ($t_{1gram}$); 2) volume of solvent required to produce 1 g of exfoliated $WS_2$ powder ($V_{1gram}$); 3) the ratio between the weight of the final graphitic material and the weight of the starting graphite flakes, defined as exfoliation yield (Y). Since each WJM pass takes ~4.5 s to process a volume of 10 mL of the $WS_2$ powder dispersion in NMP at a concentration of 10 g $L^{-1}$ without any material loss, a $t_{1gram}$ ~0.75 min, $V_{1gram} = 0.1$ L and a Y = 100% are obtained. To the best of our knowledge, WJM exfoliation outperform any other exfoliation techniques in term of these FoM.[84] By considering industrial-like material costs (*i.e.*, US$50-500/Kg for $WS_2$[85] and US$2-5/Kg for NMP,[86] which can be also recycled and re-used), WJM emerges as a cost-effective technique, resulting in a cost of high-quality few-layer flakes of $WS_2$ in the order of ~US$ 1000/Kg (neglecting extra-costs of high-purity material purchasing and instrumentation/hourly labor costs).

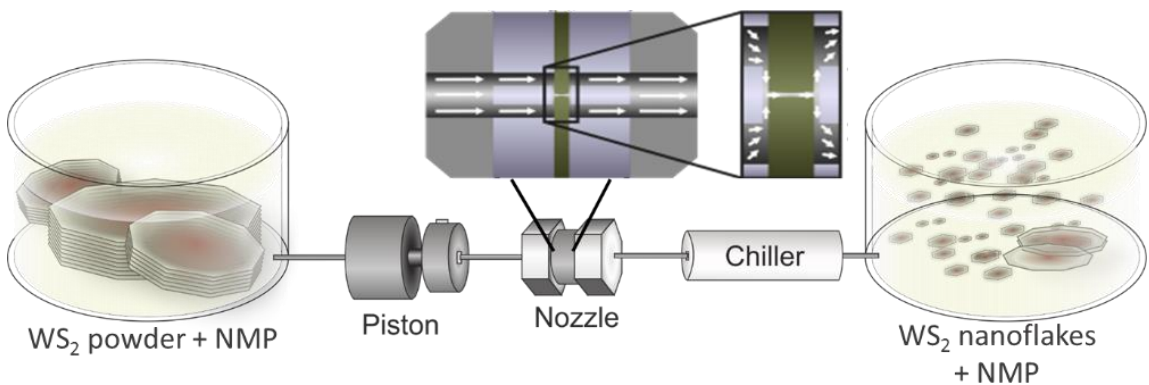

**Figure 1.** Schematic illustration of the production of $WS_2$ nanoflakes by WJM exfoliation.

The lateral size and thickness of the as-produced sample are characterized by means of transmission electron microscopy (TEM) and atomic force microscopy (AFM). The sample is composed by irregularly shaped and nm-thick flakes (**Figure 2**a,b). Statistical analysis indicates that the lateral size and thickness of the flakes approximately follow log-normal distribution peaked at ~400 nm (Figure 2c) and ~1.5 nm (Figure 2d), respectively. Optical absorption and Raman spectroscopy permit to evaluate the structural properties of the as-produced $WS_2$ flakes.[87,88,89,90] Figure 2e shows the UV-Vis absorption spectrum of the 1:100 diluted WJM-exfoliated $WS_2$ flakes dispersion. The two characteristic peaks at ~625 nm and ~525 nm arise from direct transition from the valance band of the 2H phase of $WS_2$, which is split by spin-orbit interaction,[87,88,89,90] to the conduction band at the $K$-point of the Brillouin zone, known as the $A$ and $B$ transitions.[91,92] The peaks located at ~415 and ~450 nm arise from the $C$ and $D$ interband transitions between the density of states peaks in the valence and conduction bands.[91,92,93] Figure 2f reports the Raman spectrum of the as-produced $WS_2$ flakes and the one of the $WS_2$ powder. Typically, the Raman spectrum of $WS_2$ consists mainly in three peaks: the first-order modes at the Brillouin zone center $E^1_{2g}(\Gamma)$ and $A_{1g}(\Gamma)$, which involve the in-plane displacement of W and S atoms and the out-of-plane displacement of S atoms, respectively;[88,90,94] the second order longitudinal acoustic mode at the M point, 2LA(M).[88,90,94] The $E^1_{2g}(\Gamma)$ of single-/few-layer flakes of $WS_2$, located at ~421 cm$^{-1}$, is stiffened compared to the one of bulk $WS_2$ powder, which is found at ~419 cm$^{-1}$. The blue-shift of the $E^1_{2g}(\Gamma)$ mode in $WS_2$ flakes is a consequence of the reduced dielectric screening of long-range Coulomb interaction compared to the bulk $WS_2$ powder.[88,90,94] Although the 2LA(M) overlaps the $E^1_{2g}(\Gamma)$, the multi-peak Lorentzian fitting clearly separates their individual contributions, as shown in Figure 2f. The analysis of the 2LA(M) and $A_{1g}(\Gamma)$ peak intensity ratio, *i.e.,* I(2LA)/(IA$_{1g}$), has been reported as a spectroscopic tool to assess the single-

/few-layer composition of WS$_2$ samples.[95] In our case, the spectra analysis estimates I(2LA)/(IA$_{1g}$) values ~0.3 in WS$_2$ powder and ~1.6 in WJM-exfoliated WS$_2$ flakes. These values correspond to those measured for bulk WS$_2$ (< 0.5)[95] and single-/few-layer WS$_2$ flakes (> 0.5),[95] respectively. Raman statistical analysis is reported in Supporting Information (**Figure S2**). This result, together with morphological TEM and AFM analysis, indicates that WJM exfoliation process breaks effectively the weak van der Waals force of the pristine WS$_2$ flakes without deteriorating the covalent bonds within each layer, *i.e.*, their crystal structure. X-ray photoelectron spectroscopy (XPS) measurements are carried out to ascertain the elemental composition of the WJM-exfoliated WS$_2$ flakes, *i.e.* their chemical quality. Figure 2g reports the W *4f* and W *5p* XPS spectrum of WS$_2$ flakes, together with their deconvolution. The peak at the lowest binding energy, *i.e.*, 32.9 eV, is assigned to W *4f$_{7/2}$*.[96,97] The peak at a binding energy of ~35 eV is fitted with two components. The first component peaked at 35.1 eV is assigned to W *4f$_{5/2}$* of the 2H-phase of WS$_2$, while the second one at 35.9 eV is associated to W *4f$_{7/2}$* of oxidized species (*i.e.*, WO$_3$).[97,98] Lastly, the third peak, at a binding energy of ~38 eV, is also fitted with two components. The first one at ~38.1 is assigned to W *4f$_{5/2}$* of WO$_3$,[97,98] while the second one peaked at 38.4 eV is assigned to W *5p$_{3/2}$* of WS$_2$.[96,97] Notably, the oxides-related peaks correspond to a percent content of oxide less than 20%, which means that the WJM exfoliation process does not remarkably affect the chemical quality of the WS$_2$ powder. In fact, WJM exfoliation avoids long-lasting local high-temperatures (of the order of thousands K)[99,100] and steep local heating/cooling gradients,[99] typically present in the ultrasound process,[99,100] which cause material deterioration during liquid-phase exfoliation (LPE)[101,102] of TMDs in NMP (percentage content of oxides > 40% for LPE in NMP).[103,104]

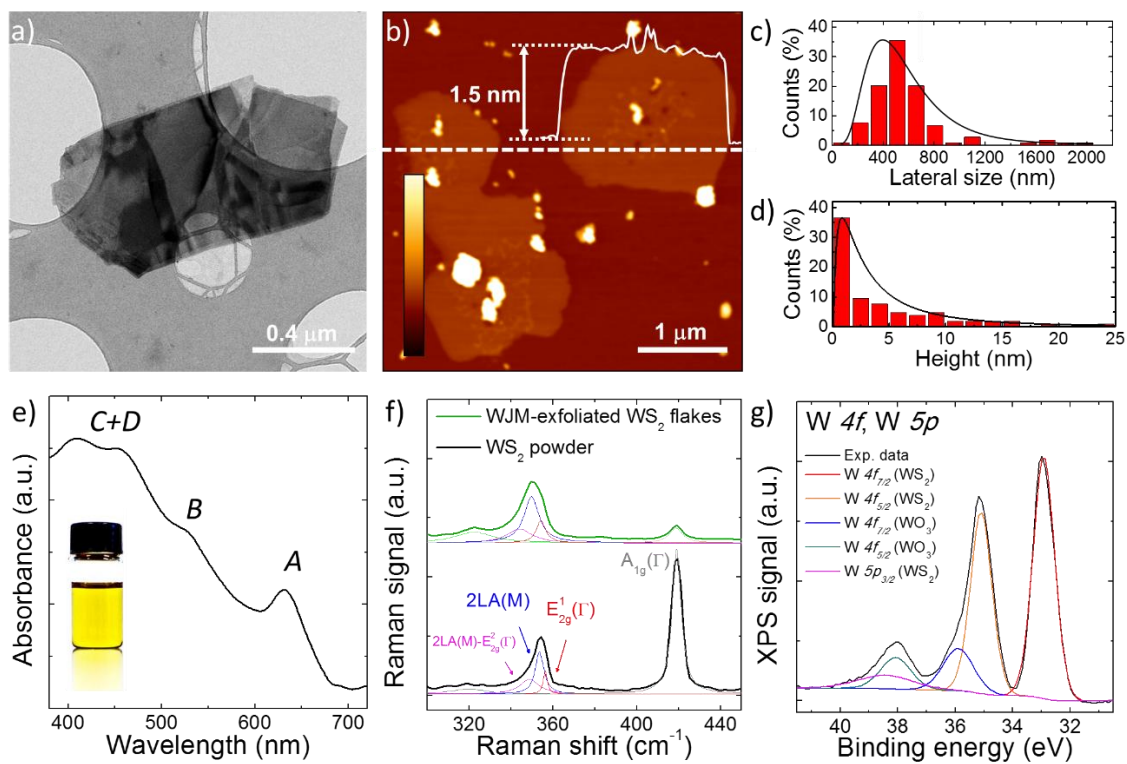

**Figure 2. Morphological, structural and chemical characterization of the WS₂ flakes produced by WJM exfoliation.** a) TEM and b) AFM images of representative WJM-exfoliated WS₂ flakes. Height profile of a representative flakes imaged in (b) is also shown (dashed white line). c) Statistical TEM analysis of the lateral dimension and d) statistical AFM analysis of the thickness of the WJM-exfoliated WS₂ flakes. e) Absorption spectrum of 1:100 diluted WJM-exfoliated WS₂ flakes dispersion in NMP, whose photograph is also shown as inset. f) Raman spectra of WS₂ powder and the WJM-exfoliated WS₂ flakes, with their multi-peak Lorentzian fitting showing the contribution of the individual modes (red line: $E^1_{2g}(\Gamma)$; grey line: $A_{1g}(\Gamma)$; blue line: 2LA(M); magenta line: 2LA(M)-$E^2_{2g}(\Gamma)$). g) W *4f* and W *5p* XPS spectrum of the WJM-exfoliated WS₂ flakes. Its deconvolution is also shown, evidencing the bands ascribed to: W *4f₇/₂* and W *4f₅/₂* of WS₂ (red and orange curves, respectively); W *4f₇/₂* and W *4f₅/₂* of WO₃ (blue and cyan curves, respectively); W *5p₃/₂* of WS₂ (magenta curve).

**Fabrication and electrochemical characterization of WS₂-DIBs.** The high-quality few-layer WS₂ flakes produced by WJM, are exploited as anode materials for the realization of WS₂/graphite DIBs, with graphite used as cathode.[26,27,28] The WS₂-anodes are prepared by coating the NMP-based slurry containing the WS₂ flakes, acetylene black and polyvinylidene difluoride (PVDF) in a weight ratio of 8:1:1 on Cu foil. Additional details of the electrode preparation and cell assembly are reported in Supporting Information. **Figure 3**a shows the first three cycles of the cyclic voltammetry (CV) analysis performed on $WS_2$ flakes electrode ($WS_2$-anode) assembled in a half-cell configuration, namely using Li foil as both counter and reference electrode. A 1 mol $L^{-1}$ $LiPF_6$ solution in ethylene carbonate (EC):ethyl methyl carbonate (EMC) (volume ratio, 1:1) ($LiPF_6$/EC:EMC) is used as electrolyte. During the first cycle, the series of three partially overlapped cathodic processes starting at ~0.8 V *vs.* $Li^+$/Li (indicated by red arrows if Figure 3a) are sequentially attributed to: 1) the solid electrolyte interface (SEI) formation at the $WS_2$ surface;[105] 2) the $Li^+$ intercalation between $WS_2$ layers; 3) the material conversion from $Li_xWS_2$ to elemental W and $Li_2S$.[59,60,61,68,69,70,71,72,73,74,78,79,80] During the first anodic scan, the peaks around 1.75 V *vs.* $Li^+$/Li and 2.5 V *vs.* $Li^+$/Li (indicated by blue arrows) are assigned to the de-lithiation of residual $Li_xWS_2$ and the de-lithiation of $Li_2S$, respectively.[59,60,61,68,69,70,71,72,73,74,78,79,80] The curve-overlapping and the unchanged positions of the redox peaks for the second and third cycles imply that the reversible electrochemical behaviour of the $WS_2$-anode is reached after the first cycle. Figure 3b shows the CV curves (measured after CV curves shown in Figure 3a) at various voltage scan rates (from 0.5 to 2 mV $s^{-1}$). The intensities of the redox peaks scale with the square root of the scan rate, as expected for diffusion-limited faradaic processes.[106]

The CD profiles, as obtained from galvanostatic cycling, well agree with peaks position and intensity as obtained from CV analysis (Figure 3c). The sloping profile of the first discharge occurring between 1.5 and 1.0 V *vs.* $Li^+$/Li is attributed to the solid electrolyte interface (SEI) formation at the $WS_2$ surface[105] as well as the $Li^+$ insertion between $WS_2$ layers.[59,60,61,68,69,70,71,72,73,74,78,79,80] The absence of a net plateau referred to this first $Li^+$ insertion might be connected to the not-well defined and separated staging insertion in few-layers $WS_2$ flakes, contrarily to what is observed for bulk $WS_2$.[107] The following two plateaus, set approximatively at 1.0 and 0.75 V *vs.* $Li^+/Li$, clearly stand for two-phase mechanisms and in particular can be attributed to further $Li^+$ insertion in $Li_xWS_2$ lattice accompanied by conversion of Li-rich $WS_2$ phase into metallic W and $Li_2S$.[59,60,61,68,69,70,71,72,73,74,78,79,80] During the subsequent charge process, the charging curve shows sloping profiles, with two plateaus at ~1.85 and ~2.3 V *vs.* $Li^+/Li$. These features are due to the de-lithiation of residual $WS_2$ and the de-lithiation of $Li_2S$, respectively.[59,60,61,68,69,70,71,72,73,74,78,79,80] The $WS_2$-anode delivers an initial discharge specific capacity of 577 mAh $g^{-1}$ and a charge specific capacity of 457 mAh $g^{-1}$ at a current rate of 0.1 A $g^{-1}$. These CD specific capacities correspond to a Coulombic efficiency of 80.3%, a value which is similar to the ones typically observed in the first CD cycles of TMD-based anodes in LIBs.[59,60,61,68,69,70,71,72,73,74,78,79,80] The irreversible initial capacity loss mainly results from the material conversion from $WS_2$ to elemental W and $LiS_2$, as well the SEI formation.[59,60,61,68,69,70,71,72,73,74,78,79,80] During the subsequent CD cycles at increasing current rates (from 0.2 A $g^{-1}$ to 1 A $g^{-1}$), the $WS_2$-anode exhibits discharge specific capacities of 438, 421 and 403 mAh $g^{-1}$ at 0.2, 0.3 and 0.4 A $g^{-1}$, respectively, still delivering a discharge specific capacity of 295 and 169 mAh $g^{-1}$ at 0.8 and 1 A $g^{-1}$, respectively. The Coulombic efficiency progressively approach 100% during the various CD cycles. This indicates that a reversible electrochemical behavior of the $WS_2$-anode is progressively reached over CD cycling, in agreement with the CV analysis (Figure 3a). It is worth noting that both soluble high-order polysulfides ($Li_2S_n$, $3 \leq n \leq 8$) and insoluble sulfides $Li_2S_2/Li_2S$ are formed after the first lithiation cycles of $WS_2$. Therefore, the subsequent CD loops of such species resemble the redox chemistry of cathode in Li-S batteries.[22,23] In such systems, the dissolved $Li_2S_n$ shuttle between the anode and cathode during the charge/discharge processes involving side reduction reactions with lithium anode and re-oxidation reactions at the cathode. These issues can negatively affect both the use of active material and the cycling stability.[108,109] Composite materials, such as S–carbon[110,111,112] and S–conductive polymer[113,114,115] composites, have been developed to constrain the electrochemical reactions of Li-S chemistry inside the corresponding nanoporous electrodes by favoring the adsorption of the $Li_2S_n$ on their surface.[108,109,116] In this context, the $W/WS_2$ network,[68,69,70,71,72,73,74,78,79,80] and the solid components of the S-species[117,118] can synergistically limit the dissolutions of the soluble polysulfides. This hypothesis agree with theoretical studies on the anchoring effects of various layered-structured materials, including TMDs, for $Li_2S_n$.[119] Furthermore, the SEI formed during the first lithiation could also prevent polysulfide diffusion effects,[120,121] limiting the anode capacity fading. Lastly, it has been proved in Li-S batteries that sulfur-deficient TMD nanoflakes catalyze the conversion of $Li_2S_n$ to $LiS_2$ (during discharge) and to elemental S (during recharge),[122] avoiding the accumulation of chemically reactive species towards carbonate electrolyte.[122,86] **Figure S3** shows the cycling behavior of the $WS_2$-anode over more than 100 CD cycles, indicating a satisfactory discharge capacity retention (97.1% after 100 CD cycles). Figure 3d shows the first cycles of the CV curves of the cathode based on graphite (graphite-cathode), as measured in the same half-cell configuration adopted previously for $WS_2$-anode. The several oxidation peaks in the potential between 4.7-5.3 V vs. $Li^+/Li$ (indicated by blue arrows) corresponds to the staged phase transformation of graphite due to $PF_6^-$ intercalation, in agreement with previous studies.[26,27,28] During the cathodic scan (Figure 3d), reduction peaks (indicated by red arrows) are observed due to the de-intercalation of $PF_6^-$ from the graphite.[26,27,28] Noteworthy, irreversible faradaic processes, due to the electrolyte oxidation, are not occurring at potentials above 4.5 V *vs.* $Li^+/Li$. In fact, **Figure S4**a shows that the electrochemical stability potential window of $LiPF_6$/EC:EMC is up to 5.5 V *vs.* $Li^+/Li$. Therefore, $LiPF_6$/EC:EMC can be considered as a reliable electrolyte formulation for DIBs, in agreement with previous studies.[29,123,124] Figure 3e shows the CV curves at various voltage scan rates (from 0.5 to 2 mV s$^{-1}$). The intensities of the redox peaks of the CV curve increase with increasing the scan rate, while retaining their shape. The intensities of the redox peaks scale with the square root of the scan rate, as expected for diffusion-limited faradaic processes.[106] Figure 3e reports the CD curves of the graphite-cathode, showing that the PF$_6^-$ intercalation mainly occur above 4.5 V *vs.* Li$^+$/Li. The specific discharge capacities of graphite-cathode are 95, 84, 80 and 72 mA h g$^{-1}$ at 0.2, 0.4 and 0.6 and 1.0 A g$^{-1}$, respectively, with Coulombic efficiency of ~96% (except for the first cycle, ~89%). Such Coulombic efficiency values inferior to 100% are attributed to the progressive graphite delamination process occurring after repetitive insertion of PF$_6^-$ anions, which can induce mechanical stresses.[26,27,28] However, cycling stability measurements (Figure S4b) shows that the graphite cathodes can retain 95.6% and 84.1% of the initial discharge capacity after 100 and 500 CD cycles, respectively, which is promising for the development of stable DIBs.[29-37]

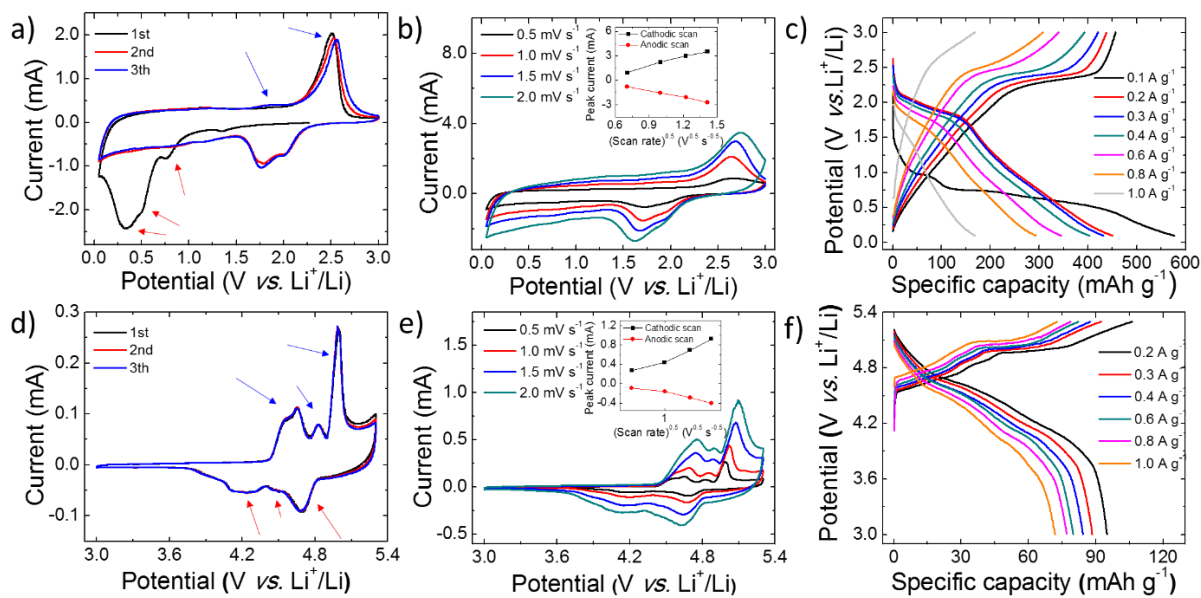

**Figure 3. Electrochemical characterization of the WS$_2$-anode and the graphite-cathode in half-cell configuration in LiPF$_6$/EC:EMC electrolyte.** a) Cyclic Voltammetry (CV) of WS$_2$-anode at 1 mV s$^{-1}$. b) CV curves of WS$_2$-anode at various voltage scan rates. c) Charge-Discharge (CD) curves of WS$_2$-anode at various current rates. d) CV cycles of graphite electrode at 1 mV s$^{-1}$. e) CV curves of graphite-cathode at various voltage scan rate. f) CD curves of graphite-cathode at various current rates. Red and blue arrows indicate reduction and oxidation stages, respectively, of WS$_2$-anode (panel a) and graphite-cathode (panel d).

The WS$_2$-graphite full-cell DIB is shown in **Figure 4**. The WS$_2$-graphite DIB working mechanism is illustrated in **Figure 4**a, being based on the combination of previously characterized WS$_2$-anode and graphite-cathode in the LiPF$_6$/EC:EMC electrolyte.

Briefly, during the charging process, the Li$^+$ and PF$_6^-$ intercalate into the WS$_2$-anode and graphite-cathode, respectively. The electrons move in parallel to the Li$^+$ ion to the negative electrode *via* the external circuit. During discharging process, both Li$^+$ and PF$_6^-$ are released back from the electrodes into the electrolyte. The operating voltage of WS$_2$-graphite DIBs ranges from 0 to 4 V with an average value of ~2.4 V, as indicated from its CV curve at voltage scan rate of 1 mV s$^{-1}$ (Figure 4b). Figure 4c shows the CD curves of our WS$_2$-graphite DIB at various current rates.

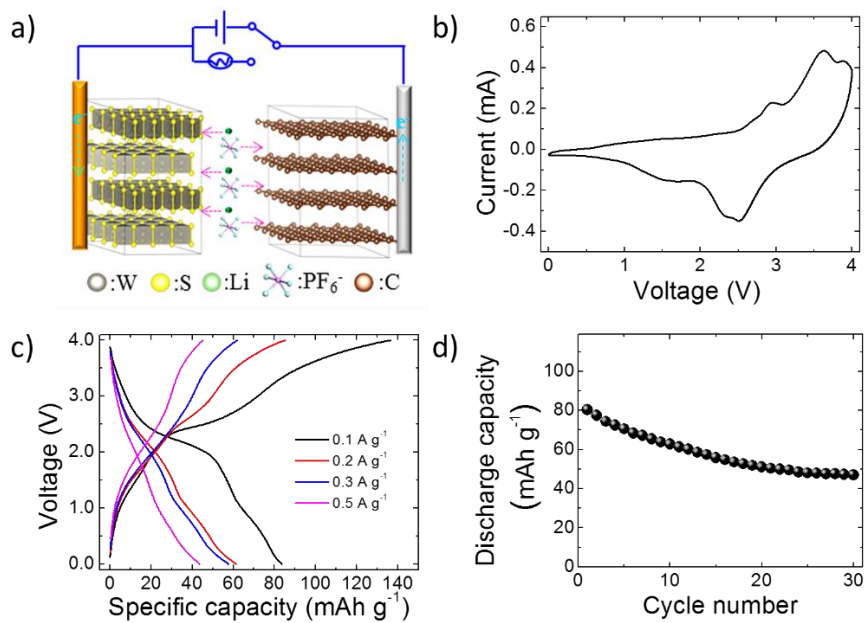

**Figure 4. Electrochemical characterization of the WS₂-graphite DIB.** a) Schematic illustration of the WS₂-graphite DIB and its working mechanism. b) CV cycles of WS₂-graphite DIB at 1 mV s⁻¹. c) CD curves WS₂-graphite DIB at various current rates. d) Discharge capacity retention of WS₂-graphite DIB over 30 CD cycles at current rate of 0.1 A g⁻¹.

The specific discharge capacities are 83, 61, 57 and 44 mA h g$^{-1}$ at 0.1, 0.2, 0.3 and 0.5 A g$^{-1}$, respectively (Figure 4c). The low Coulombic efficiency (~65%) at the initial cycles is attributed to the formation of SEI layer at the WS₂-anode side, the WS₂-anode decomposition in W and Li₂S, as well as graphite delamination process, in agreement with half-cells characterization (Figure 3). After 30 cycles, the DIB can maintain 47 mA h g$^{-1}$ with a capacity retention of 58.53 % (Figure 4d). These values set the basis for further optimization of this class of novel devices.

In order to identify possible routes for further improvement of the electrochemical perfomance of the WS₂-graphite DIBs, metallic 1T-WS₂ (distorted octahedral phase)[50] flakes were exploited as the anode material. The metallic 1T-WS₂ are produced by chemical exfoliation (*i.e.*, Li-intercalation method)[125,126,127] of WS₂ powder (see details in Experimental Section). While this production method is difficult to be scalable, this test allows to assess the impact of a higher electron conductivity of the 1T-WS₂-anode,[128,129,] compared to the one of the 2H-phase of WS₂, in the final performance of the DIBs. 1T-WS₂-anode delivers specific discharge capacities of 896, 650, 475 and 340 mAh g$^{-1}$ at 0.1, 0.3, 0.6 and 1.0 A g$^{-1}$ (**Figure S5a**), respectively. These specific discharge capacities values are higher than those obtained by using WJM-exfoliated WS₂ flakes (Figure 3c). The higher electrical conductivity of the 1T phase compared to the 2H one also improves the performance of the WS₂-graphite DIB, which achieves a specific discharge capacity of ~92 mAh g⁻¹ at 0.1 A g⁻¹ (Figure S5b). This value corresponds to a ~11% increase of the specific capacity compared to the one obtained by using WJM-exfoliated WS₂ flakes. After 10 cycles, the specific capacity is retained at 70 mAh g$^{-1}$ (Figure S5c). At higher current rates, *i.e.*, 0.2, 0.3 and 0.5 A g$^{-1}$, the specific discharge capacities are 75, 60 and 54 mAh g$^{-1}$, respectively.

**Conclusions.** In summary, our work reports the development of a novel WS$_2$-graphite dual-ion battery (DIB) combining together a conventional graphite cathode and high-capacity few-layer WS$_2$ flakes anode. The few-layer WS$_2$ flakes are produced by exploiting wet-jet milling (WJM) exfoliation. This technique allows production rate of WS$_2$ dispersion up to 8 L h$^{-1}$ at concentration of 10 g L$^{-1}$,[84] with an exfoliation yield of 100%, proving an outstanding scalability in comparison to any other techniques used for the exfoliation of transition metal dichalcogenides (TMDs).[84] The WJM-exfoliated flakes show excellent morphological, structural and chemical properties, without requiring post-production purification step, such as sedimentation-based separation,[130,102] as typically applied for 2D-material dispersion produced by liquid-phase exfoliation (LPE).[101,102] In particular, few-layer WS$_2$ flakes display lateral dimension up to ~2 µm (log-normal distribution peaking at ~400 nm), crystalline retention with respect to the native bulky powder and limited content of oxides (percentage content < 20%). When used as DIB anodes, the few-layer WS$_2$ flakes achieved specific reversible capacities of 457 and 169 mAh g$^{-1}$ at the current rates of 0.1 and 1 A g$^{-1}$, respectively. These values overcome most of the conventional DIB-anodes, mainly based on graphite and metal.[26,27, 34,48,49] In addition, the coupling of few-layer WS$_2$ flakes as anode with graphite-based cathode allows the realization of DIBs operating in the 0-4 V range, with an average value of ~2.4 V. It is worth noting that when WS$_2$-anodes are applied into lithium-ion batteries (LIBs) using commercial cathode as LiFePO$_4$ or LiCoO$_2$,[7] the operating average cell voltage is expected to be less than 1 V for LiFePO$_4$ and 1.6 V for LiCoO$_2$, as a consequence of the high operating discharge potential of WS$_2$-anode in full device (~1.8 V *vs.* Li$^+$/Li in our case, at 0.1 A $g^{-1}$). Such features currently represent the key-disadvantages of using $WS_2$,[59,61,68,72,73,107,131,132,133,134] and more in general TMDs,[135,136,137] as anode in practical LIBs, thus missing the chance of using a low-cost and high-capacity anode material. Here, we demonstrate that TMD-based anodes are promising candidate materials with high capacitance (~433 mAh $g^{-1}$ for $WS_2$ uptaking of 4 $Li^+$,[68,69,70,71,72,73,74] up to ~1675 mAh $g^{-1}$ considering the charge/discharge loops of the elemental S,[68,69,70,71,72,73,74,78,79,80] as formed after the first $WS_2$ lithiation/de-lithiation loop[68,69,70,71,72,73,74]) to be exploited in the high-voltage DIB architecture. **Figure S6** reports a sketch of representative voltammograms for $WS_2$-anode, commercial ($LiFePO_4$ and $LiCoO_2$) cathodes and graphite-cathode, which roughly estimate the resulting operating voltage for both LIBs and DIBs based on $WS_2$-anode. Finally, we have shown that the replacement of semiconducting $WS_2$, produced by WJM, with metallic $1T$-$WS_2$ flakes, increases the electrochemical performances of the DIB anodes. Although the production of $1T$-$WS_2$ is not scalable, this result suggests that further improvement of DIBs based on WJM $WS_2$ anodes to address the issue of full capacity retention over charge-discharge cycling, might require optimization of the electrical conductivity of the final composite. For example, $WS_2$ (or TMD)-carbon hybrids as anode material have been already demonstrated to provide optimum combination of energy density, cycling stability and high-rate capability in LIBs.[69,70,78,135,136,137,138,139,140] These results could be directly exploited by the DIB architecture in an attempt to improve the anode conductivity. The full potential of $WS_2$, as well as of other TMDs,[141] will be evaluated by new insight and technical progress towards commercial DIBs. Our work rationalizes the use of $WS_2$, which is representative of the entire class of the TMDs, as possible candidate for the realization of high specific capacity DIBs.

**Methods**

*Wet jet milling process.* As schematically illustrated in Figure S1, the wet-jet milling system makes use of a high-pressurized jet stream to homogenize and exfoliate the sample, *i.e.*, a layered material. More in detail, a hydraulic mechanism and a piston supply the pressure in order to direct the mixture of solvent and layered crystals into the reactor, where the exfoliation is performed. Immediately after the processing in the reactor, the sample is cooled down by means of a chiller (Figure S1a). The reactor consists in a set of five different drilled disks, which form a set of interconnected channels (Figure S1b). The configuration of the disks divides the flow in two streams (Disk A), which subsequently collide in a single point. Immediately after the collision, the flow passes through the nozzle (a perforated 0.1 mm hole, Disk B). The piston takes 10 mL of the mixture solvent/layered material from the container and triggers it towards the reactor. The process is repeated *n* times (*n* = total volume to process/piston chamber volume) until to the total volume is processed. The shear forces, the implosion of cavitation bubbles and the drastic pressure changes are the phenomena that promote the sample exfoliation in the reactor.[84] The time during which the flakes are subjected to exfoliation is less than one second,[84] compatibly with industrial volume production.

*Exfoliation of WS$_2$ powder.* WS$_2$ powder (particle size < 2 μm, 99%, Sigma Aldrich) is exfoliated by both WJM[84] and Li-intercalation method (chemical exfoliation).[125,126,127] WJM exfoliation permits to obtain WS$_2$ nanoflakes, maintaining the natural 2H phase of the native powder.[84] Briefly, 50 g of each material is dispersed in 5 L of NMP. A pressure of 250 MPa is applied in the reactor. Each piston pass process 10 mL of the WS$_2$ powder dispersion. Finally, the processed sample, *i.e.*, WJM-exfoliated WS$_2$ flakes dispersion, is collected in an end-container. Li-intercalation method is used to prepare metallic 1T-WS$_2$ flakes.[125,126,127] Experimentally, 0.3 g of WS$_2$ powder is dispersed in 4 mL of 2.0 M n-butyllithium (n-BuLi) in cyclohexane (Sigma Aldrich). The dispersion is kept stirring for 48 h at room temperature under Ar atmosphere, while Li-intercalated WS$_2$ (Li$_x$WS$_2$) is formed. The formed material is separated by filtration under Ar and subsequently washed with anhydrous hexane (Sigma Aldrich) to remove non-intercalated Li$^+$ and organic residues. The as-produced Li$_x$WS$_2$ powder is exfoliated by ultrasonication in a sonic-bath (Branson® 5800 cleaner, Branson Ultrasonics) in deionized water for 1 h. The obtained dispersion is then ultracentrifugated at 17000 $g$ (in Optima™ XE-90 ultracentrifuge, Beckman Coulter) for 20 min to remove LiOH and un-exfoliated material. Finally, the precipitate is filtered and re-dispersed in 2-propanol (Sigma Aldrich), to obtain the metallic 1T-WS$_2$ flakes dispersion. All the as-produced dispersions are dried in form of powder before the realization of the WS$_2$-anodes.

*Material characterization*

Transmission electron microscopy images are taken with a JEM 1011 (JEOL) transmission electron microscope, operating at 100 kV. Samples for the TEM measurements are prepared by drop-casting the WS$_2$ flakes dispersions onto C-coated Cu grids, rinsed with deionized water and subsequently dried under vacuum overnight. Morphological and statistical analysis is carried out by using ImageJ software (NIH) and OriginPro 9.1 software (OriginLab), respectively.

Atomic force microscopy images are taken using a Nanowizard III (JPK Instruments, Germany) mounted on an Axio Observer D1 (Carl Zeiss, Germany) inverted optical microscope. The AFM measurements are carried out by using PPP-NCHR cantilevers (Nanosensors, USA) with a nominal tip diameter of 10 nm. A drive frequency of ~295 kHz is used. Intermittent contact mode AFM images (512×512 data points) are collected by keeping the working set point above 70% of the free oscillation amplitude. The scan rate for acquisition of images is 0.7 Hz. Height profiles are processed by using the JPK Data Processing software (JPK Instruments, Germany) and the data are analysed with OriginPro 9.1 software. Statistical analysis is carried out by means of Origin 9.1 software on four different AFM images for each sample. The samples are prepared by drop-casting $WS_2$ flakes dispersions onto mica sheets (G250-1, Agar Scientific Ltd., Essex, U.K.) and dried under vacuum.

Optical absorption spectroscopy measurements are carried out using a Cary Varian 6000i UVvis-NIR spectrometer using quartz glass cuvette with a path length of 1 cm. $WS_2$ flakes dispersions obtained by WJM are diluted 1:100 with NMP before the measurements. The metallic 1T-$WS_2$ flakes dispersions are instead characterized as-produced. The corresponding solvent baselines are subtracted to the as-acquired absorption spectrum.

Raman spectroscopy measurements are carried out using a Renishaw microRaman Invia 1000 using a 50× objective, with an excitation wavelength of 532 nm and an incident power on the samples of 1 mW. For each sample, 50 spectra are collected. The samples are prepared by drop casting $WS_2$ flake dispersions onto $Si/SiO_2$ substrates and dried under vacuum. The spectra are fitted with Lorentzian functions. Statistical analysis is carried out by means of OriginPro 9.1 software.

X-ray photoelectron spectroscopy characterization is carried out on a Kratos Axis UltraDLD spectrometer, using a monochromatic Al Kα source (15 kV, 20 mA). The spectra are taken on a 300 μm × 700 μm area. Wide scans are collected with constant pass energy of 160 eV and energy step of 1 eV. High-resolution spectra are acquired at constant pass energy of 10 eV and energy step of 0.1 eV. The binding energy scale is referenced to the C 1s peak at 284.8 eV. The spectra are analysed using the CasaXPS software (version 2.3.17). The fitting of the spectra is performed by using a linear background and Voigt profiles. The samples are prepared by drop-casting $WS_2$ flakes onto Si/SiO$_2$ substrate (LDB Technologies Ltd) and dried under vacuum.

*Electrodes and DIBs assembly and electrochemical characterization*

The $WS_2$-anodes are prepared by coating a Cu foil with the NMP-based slurry containing the $WS_2$ flakes, acetylene black (Sigma Aldrich) and PVDF (Sigma Aldrich) in a weight ratio of 8:1:1, using a doctor-blade technique. The coated foils are dried and punched into circular pieces with 11 mm diameter. The obtained mass loading of $WS_2$ is 1.5 mg cm$^{-2}$. A Celgard 2730 membrane and 1 mol L$^{-1}$ LiPF$_6$ solution in EC/ EMC) (volume ratio, 1:1 (LiPF$_6$/EC:EMC) (Sigma-Aldrich) are used as the separator and the electrolyte, respectively. The graphite-cathodes are prepared by coating a stainless steel foil with a NMP-based slurry containing the graphite powder (Sigma Aldrich), acetylene black and PVDF in a weight ratio of 8:1:1. Half-cells are assembled using coin-type 2032 model cells, with Li foil as the counter and reference electrodes. The $WS_2$-graphite DIBs are assembled by using the as-prepared $WS_2$- and graphite-electrodes as anode and cathode, respectively, using coin-type 2032 model cells. The ratio between the active material mass loading of the $WS_2$-anode and that of graphite-cathode was 0.25:1, which was determined by balancing the charge store in the anode with that stored in cathode. The cells are assembled in an Ar-filled glove box with O$_2$ and H$_2$O content below 0.1 ppm. Electrochemical measurements of the $WS_2$-anodes and graphite-cathodes are performed using half-cell configuration (Li foil as both counter and reference electrode). Both the half-cells and the $WS_2$-graphite DIBs are measured at room temperature. The CV curves and charge/discharge tests are carried out by using CHI440B electrochemical working station (Chenhua, P.R. China) and a Land tester (CT2001A), respectively.

## AUTHOR INFORMATION

### Corresponding Author

* E-mail: francesco.bonaccorso@iit.it

## AUTHOR CONTRIBUTIONS

The manuscript was written through contributions of all authors. All authors have given approval to the final version of the manuscript.

## FUNDING SOURCES

This project has received funding from the European Union's Horizon 2020 research and innovation program under grant agreement No. 785219—GrapheneCore2. This work was supported by the German Research Foundation (DFG) within the Cluster of Excellence 'Center for Advancing Electronics Dresden' (cfaed).

## ACKNOWLEDGMENT

We thank the Electron Microscopy facility – Istituto Italiano di Tecnologia for support in TEM data acquisition.

# Supporting information

**Wet-jet milling apparatus**

**Figure S1** reports the schematic illustration of the wet-jet milling (WJM) apparatus. Figure S1a shows the flow of the WJM-exfoliation process, while Figure S1b illustrates a close-up view of the reactor. The detailed description of the both panels is reported in the main text (Methods section).

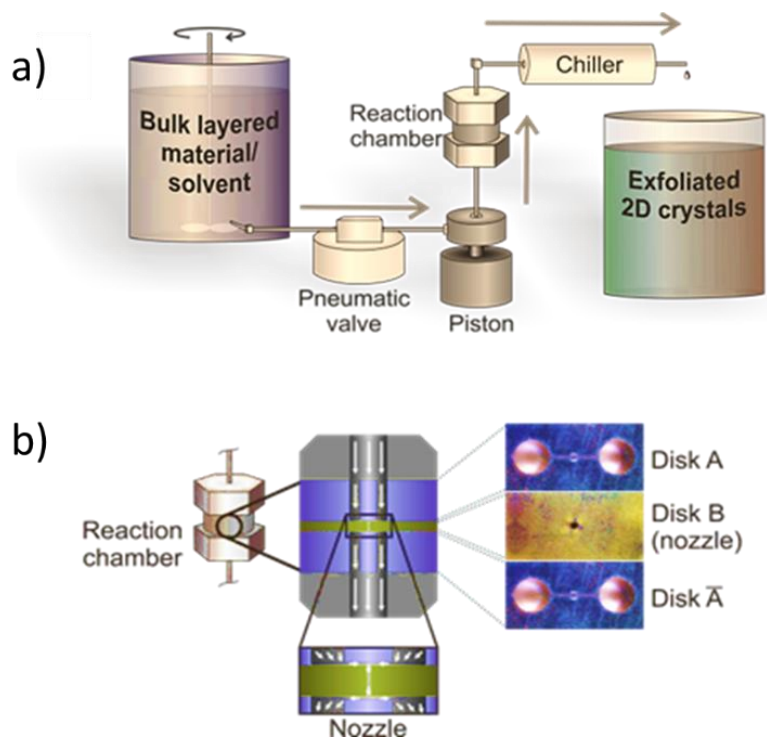

**Figure S1.** a) Schematic illustration of the wet-jet milling (WJM) apparatus. The arrows indicate the flow of the WJM-exfoliation process. b) Close-up view of the reactor. The zoomed image in (b) shows the channels configuration and the disks arrangement. The fluid path is indicated by the white arrows. On the right side, a top view of the holes and channels on each disk. The disks A and Ā have two holes of 1 mm in diameter, separated by a distance of 2.3 mm from centre to centre and joined by a half-cylinder channel of 0.3 mm in diameter. The thickness of the A and Ā disk is 4 mm. The disk B consists of a 0.10 mm nozzle and it is the core of the system. The thickness of the B disk is 0.95 mm.

**Raman statistical analysis of the WS$_2$ powder and the wet-jet milling-exfoliated WS$_2$ flakes**

**Figure S2** shows the Raman statistical analysis of the WS$_2$ powder and the wet-jet-milling (WJM)-exfoliated WS$_2$ flakes. These data complement the Raman analysis discussed in the main text (Figure 2e).

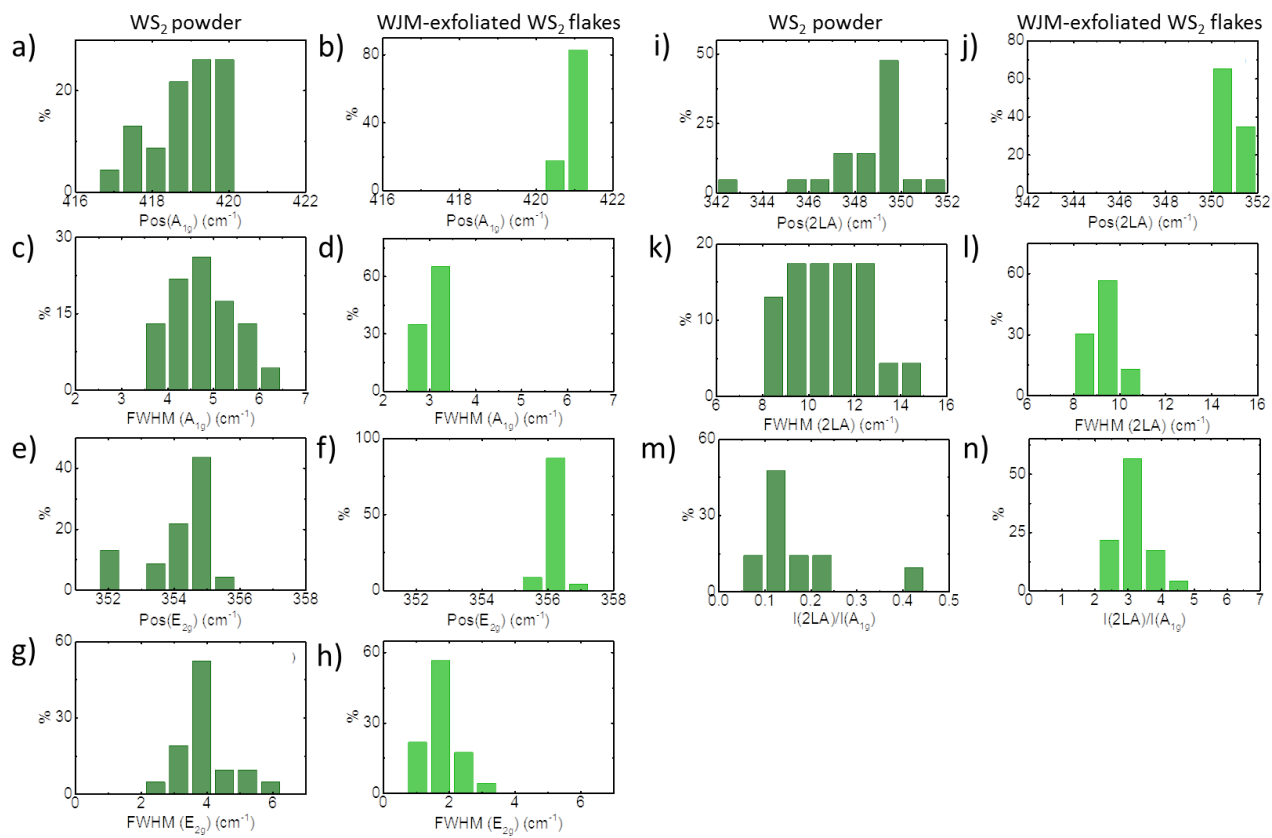

**Figure S2.** Raman statistical analysis of the WS$_2$ powder and the wet-jet milling (WTJ)-exfoliated WS$_2$ flakes. a,b) Peak position of the A$_{1g}$(Γ) (Pos(A$_{1g}$)); c,d) Full width at half maximum of the A$_{1g}$(Γ) (FWHM(A$_{1g}$)); e,f) Peak position of the E$^1_{2g}$(Γ) (Pos(E$_{2g}$)); g,h) Full width at half maximum of the E$^1_{2g}$(Γ) (FWHM(E$_{2g}$)); i,j) Peak position of the 2LA(M) (Pos(2LA)); k,l) Full width at half maximum of the 2LA(M) (FWHM(2LA)); m,n) the ratio of the intensity of the peak 2LA(M) and A$_{1g}$(Γ) (I(2LA)/(IA$_{1g}$)).

**Cycling stability of the WS₂-anode**

**Figure S3** shows the cycling behavior of the WS₂-anode over more than 100 cycles. The reversible capacity retained 97.1% of its initial value after 100 CD cycles.

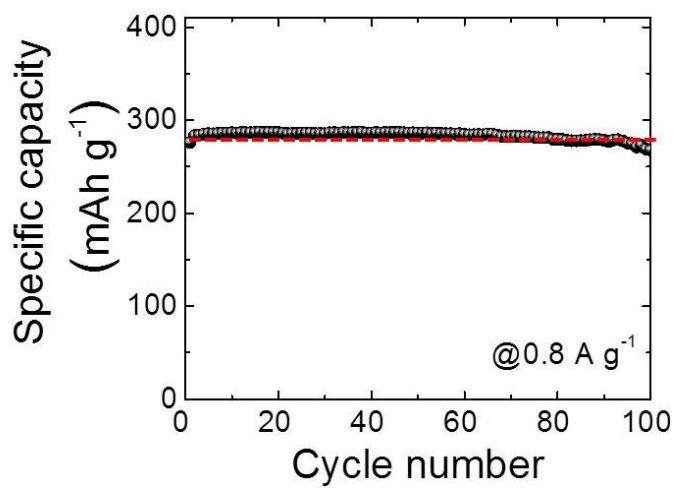

**Figure S3.** Specific capacity retention of the WS$_2$-anode over 110 CD cycles at current rate of 0.8 A g$^{-1}$.

**Stability potential window of electrolyte and cycling stability of the graphite-cathode**

**Figure S4**a shows that the electrochemical stability potential window of 1 mol L$^{-1}$ LiPF$_6$ solution

in ethylene carbonate (EC):ethyl methyl carbonate (EMC) (LiPF$_6$/EC:EMC) in comparison to that of a typical ester-based electrolyte, *i.e.*, 1 mol L$^{-1}$ LiPF$_6$ solution in 1,2-dimethoxyethane (DME):1,3-dioxolane (DOL) (volume ratio, 1:1) (LiPF$_6$/DME:DOL).

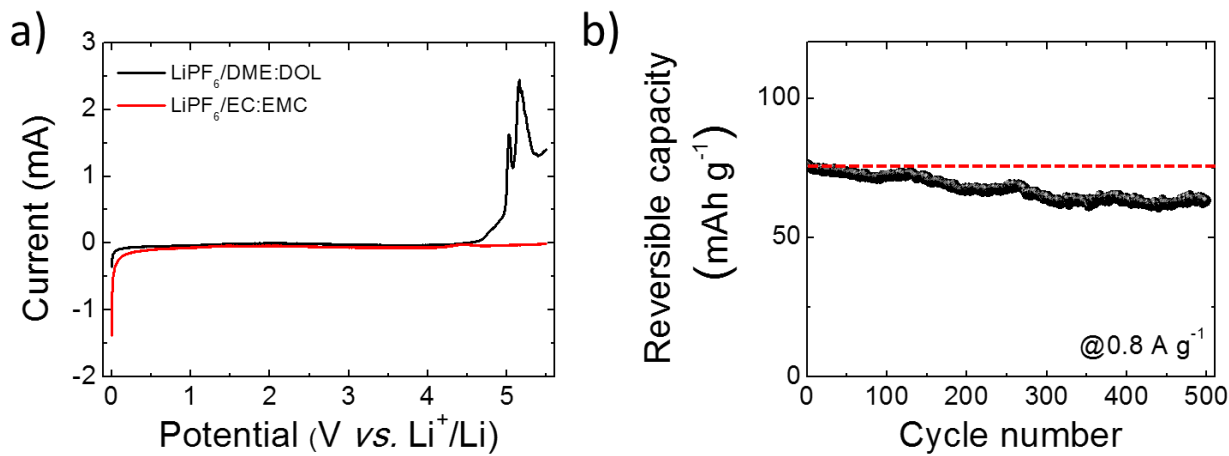

**Figure S4.** a) Linear sweep voltammograms at a scan rate of 2 mV s$^{-1}$ of a stainless steel electrode in LiPF$_6$/EC:EMC (red line) and LiPF$_6$/DME:DOL (black line). b) Reversible capacity retention of the graphite-cathode over 110 CD cycles at current rate of 0.8 A g$^{-1}$.

Clearly, only LiPF$_6$/EC:EMC can safely reach potential above 4.5 V *vs.* Li$^+$/Li (up to 5.5 V *vs.* Li$^+$/Li), where PF$_6^-$-intercalation of graphite also occurs (see main text, Figure 3d-e). Figure S4b shows the cycling behavior of the graphite-cathode over 500 cycles. The reversible capacity retained 95.6% and 84.1% of its initial value after 100 and 500 CD cycles, respectively.

**Electrochemical characterization of WS$_2$-anode and WS$_2$-graphite dual-ion battery using metallic 1T-WS$_2$ flakes**

**Figure S5** reports the electrochemical characterization of the WS₂-anode and WS₂-graphite dual-ion battery (DIB) using metallic 1T-WS₂ flakes produced by chemical exfoliation (*i.e.*, Li-intercalation method) of the WS₂ powder.

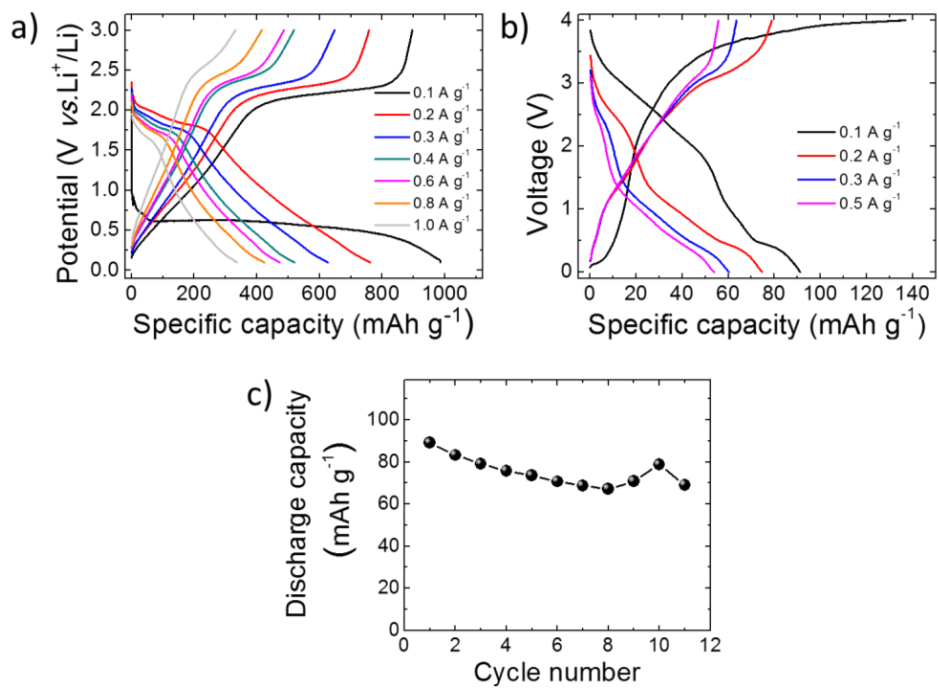

**Figure S5.** Electrochemical characterization of the WS₂-anode and the WS₂-graphite DIB using metallic 1T-WS₂ flakes. a) CD curves of WS₂-anode in LiPF₆/EC:EMC at different current rates. b) CD curves WS₂-graphite DIB at different current rates. c) Discharge capacity retention over 11 CD cycles.

LiPF₆ solution in ethylene carbonate (EC)/ethyl methyl carbonate (EMC) (volume ratio, 1:1) (LiPF₆/EC:EMC) is used as electrolyte. As shown in Figure S5a, the WS₂-anode delivers specific discharge capacities of 896, 650, 475 and 340 mAh g⁻¹ at 0.1, 0.3, 0.6 and 1.0 A g⁻¹, respectively. The WS₂-graphite DIB achieves a specific discharge capacity of ~92 mAh g⁻¹ at 0.1 A g⁻¹ (Figure S5b), which correspond to a ~11% increase of the specific capacity obtained by using WJM-exfoliated WS₂ flakes. At higher current rates, *i.e.*, 0.2, 0.3 and 0.5 A g⁻¹, the specific discharge capacities are 75, 60 and 54 mAh g$^{-1}$, respectively. After 10 cycles at 0.1 A g$^{-1}$, the specific capacity is retained at 70 mAh g$^{-1}$ (Figure S5c).

**Additional considerations on the use of WS$_2$-anodes in LIBs and DIBs**

**Figure S6** reports a sketch of representative voltammograms for WS$_2$-anode, commercial (LiFePO$_4$ and LiCoO$_2$) cathodes and graphite-cathode, estimating the resulting operating voltage for both WS$_2$-based LIBs and DIBs.

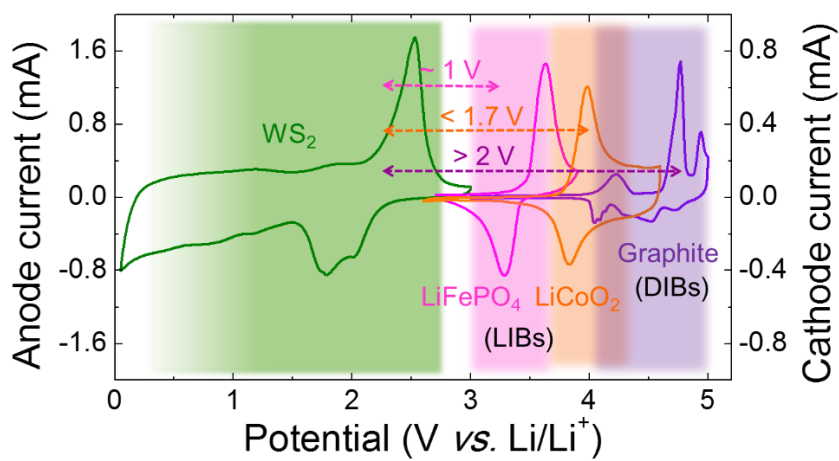

**Figure S6.** Sketch of representative voltammograms for WS$_2$-anode, commercial (LiFePO$_4$ and LiCoO$_2$) cathodes and graphite-cathode with same active material mass loading.